\documentclass{sig-alternate}

\usepackage{subfig}
\usepackage{color}
\usepackage{comment}

\usepackage[utf8]{inputenc}
\usepackage{amsmath}
\usepackage{url}
\usepackage{footnote}

\usepackage{hyperref}
\hypersetup{
    pdftitle={Mining cross-cultural relations from Wikipedia}
}

\title{Mining cross-cultural relations from Wikipedia - \\A study of 31 European food cultures}

\numberofauthors{4}
\author{
    \alignauthor
    Paul Laufer \\
    \affaddr{Graz University of Technology} \\
    \affaddr{Graz, Austria} \\
    \email{paul@laufer.at}
    \alignauthor
    Claudia Wagner \\
    \affaddr{GESIS \& U. of Koblenz} \\
    \affaddr{Cologne, Germany} \\
    \email{claudia.wagner@gesis.org}
    \and
    \alignauthor
    Fabian Flöck\\ 
    \affaddr{GESIS} \\
    \affaddr{Cologne, Germany} \\
    \email{fabian.floeck@gesis.org}
    \alignauthor
    Markus Strohmaier \\
    \affaddr{GESIS \& U. of Koblenz} \\
    \affaddr{Cologne, Germany} \\
    \email{markus.strohmaier@gesis.org}
}

\begin{document}
\maketitle



\begin{abstract}


For many people, Wikipedia represents one of the primary sources of knowledge about foreign cultures. Yet, different Wikipedia language editions offer different descriptions of cultural practices. Unveiling diverging representations of cultures provides an important insight, since they may foster the formation of cross-cultural stereotypes, misunderstandings and potentially even conflict. In this work, we explore to what extent the descriptions of cultural practices in various European language editions of Wikipedia differ on the example of culinary practices and propose an approach to mine cultural relations between different language communities trough their description of and interest in their own and other communities' food culture. We assess the validity of the extracted relations using 1) various external reference data sources (i.e., the European Social Survey, migration statistics), 2) crowdsourcing methods and 3) simulations. 
\end{abstract}

\section{Introduction}

Pierre Bourdieu emphasizes the importance of taste and related practices for analyzing culture since social groups often differentiate themselves from others via these practices \cite{Bourdieu1984}. In this work we focus on culinary practices (i.e., cultural practices of preparing and consuming food) since previous research suggests that food and culture are deeply connected \cite{Calvo1982,Talhelm2014}.


For example, a Wikipedia article on ``French cuisine'' found on the Romanian-language edition might surprise a French national when translated into her mother tongue. Unlike the French ``original'', only a brief mention of French wines exists and only a very short paragraph on croissants and pastries; but, on the other hand, it features a section on fois gras and lamb dishes so extensive that the French language pendant pales in comparison.
This might suggest that the editor community of the Romanian-language Wikipedia could either have a deviant mental picture of the French cuisine -- or it might estimate the priorities of Romanian-speaking readers to rather be on meat-based French delicatessen than on wine and baking goods. Further, the general interest of the Romanian-speaking readers in the French cuisine (for example measured by the number of views of the article about French cuisine in the Romanian language edition) might serve to potentially displease any Francophile, since the Romanian speaking community might show notably less interest in the French kitchen than in the Russian or Hungarian one. 
This hypothetical scenario serves as an example for numerous similar real-world cases (which cannot only be found in the culinary domain but also, e.g., in artistic domains) where the perception of and interest in a cultural practice differs widely over the language editions of Wikipedia, arguably the world's most culturally diverse encyclopedia when it comes to editors and readers. 

Since the percentage of people who look up information on Wikipedia is steadily increasing,\footnote{\url{http://www.pewinternet.org/2011/01/13/wikipedia-past-and-present/}} the public is likely to be significantly affected by relying on biased 
information retrieved from articles. 
However,  biased information and cultural misreading is almost unavoidable since people can only understand foreign cultures through their own ethnic culture's lenses. 
In this light, unveiling diverging (mutual) representations of cultures on Wikipedia is important, since they may explain, and even foster, the formation of cross-cultural stereotypes, misunderstandings and even conflict. 
On the other hand, cultural similarities and frequent exposure to different cultures may promote cross-cultural understanding; which in turn may lead to positive affinities between communities which may even affect the development of cross-cultural relations.
In this paper, we are interested in elucidating affinities, similarities and understanding between cultural communities by exploring the collective description and popularity of \emph{cultural practices} \cite{Bourdieu1977} within and across different language communities on Wikipedia.

\smallskip
\noindent

\textbf{Contributions and Findings.} 
In this work we demonstrate how the description of and the interest in each others food culture differ widely among European languages and propose a computational approach for mining and assessing relations between language communities on Wiki\-pedia along three dimensions (\emph{cultural understanding}, \emph{cultural similarity} and \emph{cultural affinity}).  
We evaluate the utility of our approach by assessing the plausibility of the results via (i) comparing them with external data that reveal information about relations between language communities in Europe, more precisely shared values and beliefs according to the European Social Survey and migration statistics, (ii) crowdsourcing methods and (iii) simulations.
The evaluation highlights the general applicability of our technique for cultural relation mining, while uncovering potential for including further cultural practices (e.g., literature, movies, music) to infer cultural relations beyond the culinary domain.  
Our results reveal that shared internal states such as beliefs and values that may define a culture according to Hofstede and Alavi et al.\cite{hofstede1980cultures,Alavi2004} are positively correlated with shared practices that can also be used to define a culture as suggested by Bourdieu \cite{Bourdieu1984}; the advantage of shared practices is that they can often be observed directly and are often well documented, while shared values and beliefs may only be observed indirectly if they manifest in shared behavior.
Lastly, through application of our newly introduced methodical toolset, we gain insights into patterns of cultural relations and factors that may be related to them: for instance, as suggested by Tobler's first law of geography \cite{Tobler1970}, we find that neighbouring countries tend to have more similar cultural practices than more distant countries and that cultural understanding can in part be explained by the global importance of a food culture and by migration.




\section{Related work}

Previous research acknowledged the fact that interesting differences exist in different language editions of Wikipedia \cite{hecht2009measuring,filatova2009multilingual,overell2011world,hecht2010tower,nemoto2011cultural}. Systems like Omnipedia \cite{bao2012omnipedia} or Manypedia \cite{massa2013manypedia} aim to allow users to compare and browse the different language-specific views which are inherent to Wikipedia.
For example, Hecht and Gergle \cite{hecht2010tower} showed that the diversity of information across Wikipedia language editions is much greater than initially estimated by literature and that only one-tenth of one percent is compromised of common concepts (i.e., articles about the same topic). 
In \cite{filatova2009multilingual} the authors analyze the distribution of information describing a single concept across multiple language editions of Wikipedia and find that no facts describing a concept in two different languages were contradictory, but that different language edition focused on different information. 
Previous research also found that the strongest predictor of similarity between two language editions of Wikipedia is size \cite{Warncke-Wang2012}, which suggests that one needs to be careful when using the concept similarity across different language editions as proxy for cultural similarity. 
However, our results show that inferring cultural similarity across language editions is possible if we restrict our analysis to a meaningful sub-sample of articles that belong to a domain that is related to culture. 
Our work goes beyond previous work by exploring how a meaningful sub-sample of Wikipedia articles about cultural practices are presented and consumed (i.e., viewing patterns) by the wider public and to what extent different factors may explain the cross-cultural relations which we observe on Wikipedia.

Callahan and Herring \cite{callahan2011cultural} found culturally biased differences for both, the extent and the concepts, with which the persons were described on the English and Polish Wikipedia. 
Also the editing behavior of the Wikipedia community reveals cultural differences. For example, countries with a lower Human Development Index (HDI) such as Russia or Poland show less interest in editing and maintaining Wikipedia than more developed countries such as Denmark or Germany \cite{rask2008reach}. 

Recent research in the field of computational social science has revealed the potential of non-reactive methods that are based on the analysis of large amounts of observational data for studying cultures. Examples include studies of cultural trends in literature \cite{Michel2011}, activities of Twitter users  \cite{gavilanes2013cultural}, instant messaging \cite{Kayan2006}, Flickr image tagging \cite{yanai2009detecting}, Foursquare checkins \cite{silva2014drink} or Eurovision songcontest voting data \cite{garcia2013measuring}. 
This research nicely shows that values and preferences of different cultures manifest to a certain extent in the online behavior of people and its observable outcome. 
Unlike previous work we exploit the collectively generated descriptions of culinary practices on different language versions of Wikipedia since each language community may perceive and document their own and other cultural practices through their particular cultural lenses. In addition to the content we exploit the view statistics to approximate the interest of different language communities in their own and foreign cultural practices.

\section{Methods \& Datasets}
\label{sec:approach}




Though multiple divergent definitions and measures of culture have been proposed in previous work (see \cite{Leidner2006} for an excellent overview), many scholars agree that more observable aspects of culture (e.g., norms, practices or myths) and less observable aspects of culture (e.g., beliefs or values) exist (see e.g.,  \cite{Jermier1991,Schein1985}). 
In ``Theory of Practice'' \cite{Bourdieu1977} and ``La distinction'' \cite{Bourdieu1984} Pierre Bourdieu emphasizes the importance of taste and related practices that people use to differentiate themselves from others. He argues, e.g.,~that the tastes in food, music or art and the related practices which one can observe (i.e., what people tend to eat or listen to) are indicators of a social class. 


In this work we adapt his view on culture and present a computational approach for analyzing cultural practices of distinct communities.
We use language as a proxy for cultural communities since language is closely linked to both national and cultural boundaries; it facilitates the development of a  common identity as is illustrated by the fact that the most obvious subnational divisions of cultural groups are found between language groups in multi-lingual societies such as Belgium or Canada \cite{West2004}.

\subsection{Cross-Cultural Relation Mining}
In the following, we present our approach for Cross-Cultural Relation Mining (CCRM) that looks at three distinct but interdependent dimensions of cultural relations:

\smallskip
\noindent
\textbf{Cultural Similarity.} We assess cultural similarity between two communities A and B by comparing the Jaccard Similarity\footnote{Other similarity measures might be suited as well and we plan to explore those in future work.} $\frac{F_A \cap F_B|}{|F_A \cup F_B|}$ between the sets of concepts A and B that are used to describe the cultural practice of their corresponding community.
Concepts on Wikipedia are language independent and are identified via the links (to other Wikipedia pages) of articles describing cultural practices in different language editions. For instance, if the article about the French cuisine links to concepts like ``Wine'', ``Cheese'' and ``Baguette'', the Italian cuisine references concepts like ``Wine'', ``Pasta'' and ``Cheese'' and the German cuisine is described by concepts like ``Beer'', ``Sausage'' and ``Cheese'', then we can conclude that the Italian cuisine is more similar to French than to the German one.

\smallskip
\noindent
\textbf{Cultural Understanding.} To assess the cultural understanding of a community $B$ for the culture of another community $A$, we measure the Jaccard similarity between the descriptions of $A$'s cultural practices in the language of community $B$.
If community $B$ uses the same concepts to describes $A$'s cultural practice as $A$, we can conclude that  community $B$ has a ``good'' understanding of $A$'s culture, where ``good'' means ``close-to-native''. 
If, for example, the Greek cuisine is described using the same culinary concepts (e.g., ingredients and dishes) by the Italian and Greek language communities, we can conclude that the Italian community has a good understanding of the Greek food culture. 




\label{sec:methods-affinity-bias}
\smallskip
\noindent
\textbf{Cultural Affinity and Bias.} We assess the cultural affinity between a community $A$ and $B$ (or $A$'s bias for $B$) by measuring how much attention community $A$ pays to cultural practices of community $B$. Both, the content of the articles which describe the cultural practices of a community and their access volume are used as a measure of attention. 
For instance, if the Finnish Wikipedia describes the Irish cuisine in much more detail than we would expect and/or the Irish cuisine page in the Finnish Wikipedia is viewed much more often than we would expect, then this clearly indicates that the Irish cuisine is important for the Finnish Wikipedia users.
Therefore, we can conclude that they have a positive affinity towards the Irish food culture.


Since some resources might be more popular than others, we normalize the affinities between communities by the resource popularity as follows:
\begin{equation}
\textit{bias} (l,o) = \frac{\overbrace{f(l,o)}^{\text{attention twds. res. } o}}{\underbrace{\sum_{\bar{o} \in O} f(l, \bar{o})}_{\text{attention twds. all res.}}} - \underbrace{\frac{1}{|L| - 1} \sum_{\bar{l} \in L \setminus l} \frac{f(\bar{l}, o)}{\sum_{\bar{o} \in O} f(\bar{l}, \bar{o})}}_{\text{normalized rel. attention of others}}
\end{equation}

$L$ is the set of all languages, $O$ is the set of all country-specific cultural practices which we analyze. $\textit{bias}(l,o)$ defines the bias of language $l$ towards the cultural practices $o$ of one country (in our case the cuisine of a country). The first term of the equation normalizes for the size of the language community - i.e., how important they consider the cuisine of that country compared to all other cuisines. The second term normalizes for the global importance of the cuisine.
This yields to an affinity value between $-1$ and $+1$.  A positive (negative) affinity value indicates that the community pays more (less) attention to a country-specific cultural practice than we would expect.

Beside the cultural affinity between communities, we are also interested in the affinity of each community towards itself (i.e., its \emph{self-focus bias}\cite{hecht2009measuring}) and towards geographically close communities (i.e., its \emph{regional bias}).
The self-focus bias reveals the tendency of a language community (e.g., the French community) to focus on their own cultural practice (e.g., the French cuisine).
The self-focus bias is based on the previously introduced affinity measure and is defined as follows:

\begin{equation}
\textit{sfb} (l) = \left. \underbrace{\frac{1}{|O_{\text{own}}|} \sum_{o \in O_{\text{own}}} bias(l,o)}_{\text{avg. bias towards own res.}} - \underbrace{\frac{1}{|O_{\text{other}}|} \sum_{o \in O_{\text{other}}} bias(l,o)}_{\text{avg. bias towards others}} \right.
\end{equation}
That means, we calculate the difference between the average attention towards the resources relevant to a culture $l$ and the average attention towards resources not relevant to the culture ($O_{\text{other}}$). 

The regional bias of a community is defined in the same way and measures how much more attention a community attributes to their neighbours' cultural practices compared to the practices of non-neighbouring communities\footnote{The information about the country adjacency was retrieved from \url{https://github.com/P1sec/country_adjacency}, which builds upon the Correlates of War Project (\url{http://correlatesofwar.org/}) considering two countries to be adjacent if they are separated by a border or a maximum of 24 miles of water.}.



\subsection{Datasets}

\urldef{\deitalian}\url{http://de.wikipedia.org/wiki/Italienische_K%C3%BCche} 

Altogether, 27 different language editions of Wikipedia and 31 different cuisines from across Europe were analyzed, as listed in Table \ref{tab:03-01-langs-cuisines-used}. 
Using this manually created list of cuisine articles of different European communities as a seed dataset\footnote{\url{https://www.dropbox.com/s/9rq9eqxsgnggtz3/urls.txt?dl=0}}, we created the following datasets:

\begin{table}[b!]
\centering
\scriptsize
\begin{tabular}{l|l|r|p{1.5cm}|r|r}
                  &               &                   & \textbf{Related Euro.}              &                       &                       \\ 
\textbf{Language} & \textbf{LC}   & \textbf{Size}     & \textbf{Cuisines}                   & \textbf{Editors}      & \textbf{Views}        \\ \hline
Bulgarian         & bg            &           158,130 & Bulgarian                    &                 17.91 &                  4669 \\
Bosnian           & bs            &            48,761 & Bosnian                      &                 26.00 &                   853 \\
Catalan           & ca            &           422,684 & Catalan                      &                 28.58 &                  2155 \\
Czech             & cs            &           289,551 & Czech                        &                 31.54 &                 11361 \\
Danish            & da            &           186,047 & Danish                       &                 45.50 &                  3122 \\
German            & de            &         1,692,696 & Austrian and German          &                146.34 &                108501 \\
English           & en            &         4,462,417 & British, English and Irish   &                222.77 &                468724 \\
Spanish           & es            &         1,084,184 & Spanish                      &                 77.78 &                137567 \\
Estonian          & et            &           121,329 & Estonian                     &                 13.25 &                  1221 \\
Hungarian         & hu            &           256,215 & Hungarian                    &                 41.10 &                  8603 \\
Croatian          & hr            &           143,375 & Croatian                     &                 11.67 &                  1362 \\
Finnish           & fi            &           342,384 & Finnish                      &                 22.42 &                  9467 \\
French            & fr            &         1,481,635 & French                       &                 74.64 &                 66692 \\
Italian           & it            &         1,103,118 & Italian                      &                 46.23 &                 39720 \\
Lithuanian        & lt            &           163,546 & Lithuanian                   &                 18.33 &                  4202 \\
Latvian           & lv            &            52,871 & Latvian                      &                 13.67 &                  1269 \\
Dutch             & nl            &         1,763,752 & Belgian and Dutch            &                 40.88 &                 13125 \\
Norwegian         & no            &           412,649 & Norwegian                    &                 30.75 &                  2544 \\
Polish            & pl            &         1,031,851 & Polish                       &                 46.37 &                 42972 \\
Portuguese        & pt            &           821,450 & Portuguese                   &                 37.67 &                 48972 \\
Romanian          & ro            &           241,239 & Romanian                     &                 23.00 &                  3575 \\
Russian           & ru            &         1,093,578 & Russian                      &                 58.68 &                 59685 \\
Slovak            & sk            &           190,907 & Slovak                       &                 21.67 &                  2243 \\
Serbian           & sr            &           243,268 & Serbian                      &                 25.00 &                   523 \\
Swedish           & sv            &         1,612,310 & Swedish                      &                 32.76 &                 16945 \\
Turkish           & tr            &           224,742 & Turkish                      &                 41.71 &                 17674 \\
Ukrainian         & uk            &           496,343 & Ukrainian                    &                 28.66 &                  9960 \\
\end{tabular}
\caption{\textbf{Statistics of the Dataset:} Language editions of Wikipedia, their language codes, their sizes, the related European cuisines, their average number of unique editors of cuisine articles and the average monthly views of cuisine articles (as of May 2014).}
\label{tab:03-01-langs-cuisines-used}
\end{table}

\smallskip
\noindent
\textbf{Outlink dataset}
The first dataset consists of the outgoing links of all seed articles and the language independent concept of each article to which an outlink points. 
Note that we also experimented with a two-hop dataset where we extended the set of seed articles with articles to which they point and got compareable results. 

\smallskip
\noindent
\textbf{View counts dataset}
Additionally to the content of the articles, which could in theory be heavily influenced by single contributors, the view counts represent the attention different cuisines receive. 
We use the view counts for each seed article in each language edition between May 2013 and June 2014.\footnote{\url{http://stats.grok.se/}} 
 


\begin{figure}[t!]
    \centering
    \includegraphics[width=1\linewidth,trim=0 0 0 10,clip=true]{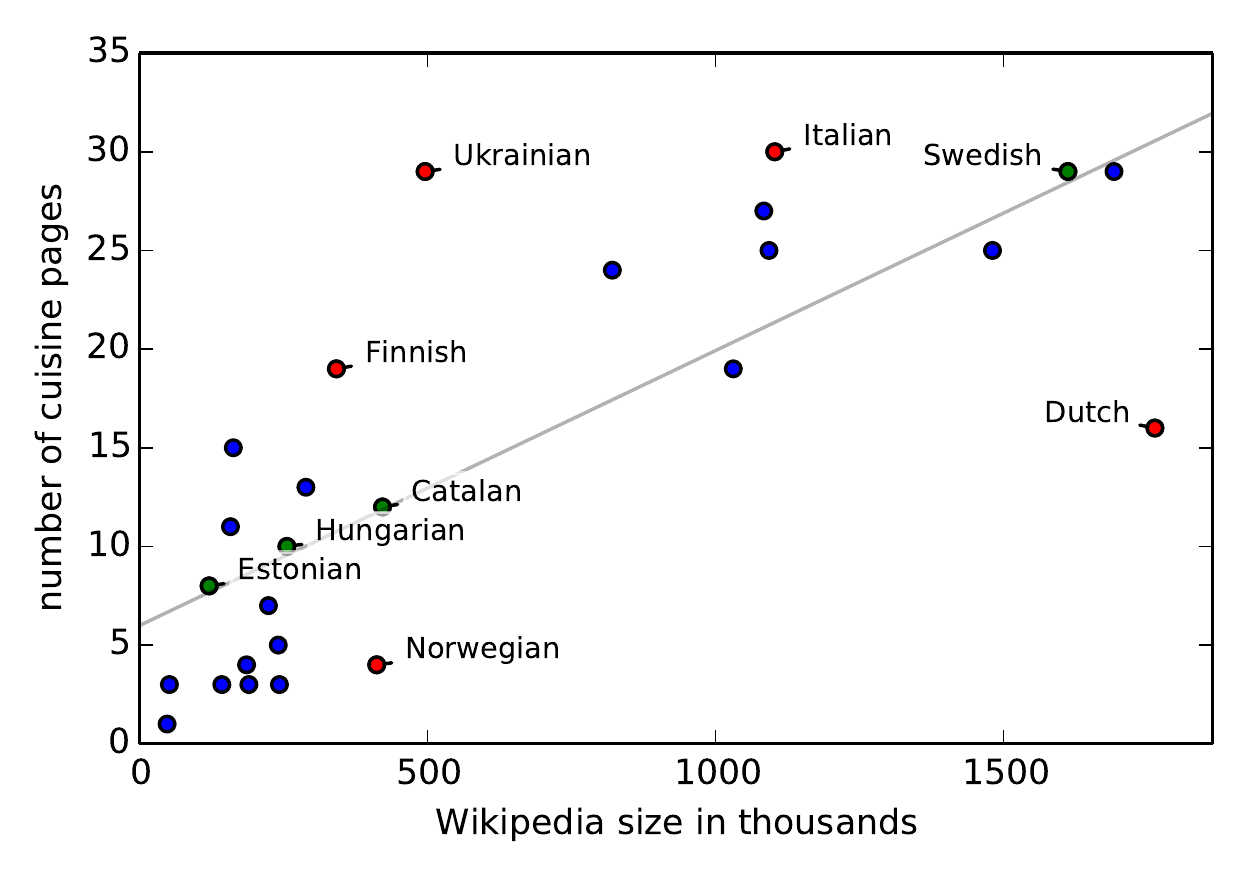}
    \caption[Size of Wikipedia language community vs. number of cuisine articles in that language edition]{\textbf{Language community's interests in different food cultures:} The plot shows the relationship between the size (in articles) of the respective language edition and the number of European cuisine articles it contains. Data points in red are furthest from the best fitting linear approximation of the relationship. Data points in green are closest to the best linear approximation. The English Wikipedia was left out, as it is considerably bigger than all other language editions.}
    \label{fig:03-01-interest-in-cuisines-per-wiki}
\end{figure}

Figure \ref{fig:03-01-interest-in-cuisines-per-wiki} shows the relation between the size of the different Wikipedia editions (i.e., their total number of pages) and the number of (European) cuisine pages that they contain. The  strong correlation (spearman correlation coefficient is $0.82$, $p \ll 0.001$) shows that larger language communities on Wikipedia indeed describe more foreign food cultures. However, some language editions tend to describe many cuisines, indicating a greater interest of their language community in food cultures, such as the Italian, Ukrainian or Finnish one, whereas others of comparable size only contain articles about very few cuisines, such as the Dutch or Norwegian Wikipedia. This suggests that the Dutch and the Norwegian language community seem to have less interest in food cultures than we would expect.

\section{Results}
In the following section we present our results on describing cultural relations on Wikipedia using the approach which we described in the previous section.

\begin{table}[b!]
\scriptsize
\centering
\begin{tabular}{l|r p{1cm} l|r}
\textbf{Cuisine}  & \textbf{Sim}  & & \textbf{Cuisine}  & \textbf{Sim}  \\ \hline
Austrian    &            +1.51    & & Irish             &  +1.50        \\
Belgian     &            +1.41    & & Italian           &  +1.23        \\
Bosnian     &            +1.26    & & Latvian           &  +2.00        \\
British     &            +1.22    & & Lithuanian        &  +1.51        \\
Bulgarian   &            +1.42    & & Norwegian         &  +1.77        \\
Catalan     &                     & & Polish            &  +1.48        \\
Croatian    &            +1.36    & & Portuguese        &  +1.72        \\
Czech       &            +1.51    & & Romanian          &  +1.42        \\
Danish      &            +1.62    & & Russian           &  +1.50        \\
Dutch       &            +1.58    & & Serbian           &               \\
English     &                     & & Slovak            &  +1.69        \\
Estonian    &            +1.70    & & Spanish           &  +1.86        \\
Finnish     &            +1.73    & & Swedish           &  +1.67        \\
French      &            +1.64    & & Turkish           &  +1.35        \\
German      &            +1.36    & & Ukrainian         &  +1.36        \\
Hungarian   &            +1.14    & &                   &               \\  
            &                     & &                   &               \\ \hline
Average     &            +1.52    & & Std.Dev           &   0.20        \\
\end{tabular}
\caption{\textbf{Cultural similarity with neighboring countries:} Ratio between the cultural similarity of neighboring countries and non-neighbouring countries with respect to their specific cuisine. A value of e.g. 2 indicates that neighboring countries are twice as similar as non-neighbouring ones. Some values are missing since we defined that a minimum of 3 neighboring countries are needed for calculating the mean.
} 
\label{tab:04-regional-similarity}
\end{table}

\subsection{Cultural Similarity}

To assess the cultural similarity between two communities with respect to their cuisines, we use the overlap of concepts that are referenced when their cuisines are described.

\subsubsection{Results}
From a global perspective the two communities which are culturally most similar with respect to their cuisine are Russia and the Ukraine, followed by Finland and Sweden and Lithuania and the Ukraine. This shows that unlike in \cite{Warncke-Wang2012} where the authors report that the size of a language edition was the strongest predictor of similarity, we do not find size-effects and a manual inspections of language pairs shows that many of them seem to be plausible and geographic close.
This raises the question to what extent cultural similarity can be explained by geographic distance. To address this question, we compare the average cultural similarity of each community with neighboring and non-neighbouring communities. 
Table  \ref{tab:04-regional-similarity} shows that geographic distances indeed can explain in part cultural similarity between communities according to their cuisines and that \emph{each cuisine is roughly $1.5$ times more similar to its neighbors than to non-neighbours} (with a standard deviation of 0.2).
This finding is in line with previous research which found that geographic distance has a large impact on culinary similarities \cite{Zhu2013} and a slight impact of similarity between language editions of Wikipedia \cite{Warncke-Wang2012}

\subsubsection{Validation}
To validate whether the cultural similarities between communities which we inferred from Wikipedia are plausible, we set up a crowd-sourcing task.
From the list of community pairs which were ranked by their cultural similarity inferred from Wikipedia, we randomly selected one pair out of the 15 most and the 15 least similar pairs and presented both pairs to the crowd workers.  We asked them to judge which pair of cuisines is more similar.\footnote{The crowsourcing platform Crowdflower was used, where we received at least 10 judgements per pair paying 4 cents per judgement. We received very positive worker feedback ($4.4$ out of $5$).} 
Out of the 225 combinations of pairs which we presented to the crowd workers, all but one where evaluated by the crowd according to our predictions as informed by the article data (199 were selected with Crowdflower confidence scores over 0.9, and 25 with scores between 0.6 and 0.9)\footnote{The score lies between 0 and 1 and describes the level of agreement between multiple contributors (weighted by the contributors’ trust scores), cf. \url{http://tinyurl.com/CFconfscore}}.
This means that for 99.56\% of all pairs, crowd workers agreed with the high similarity and the low similarity pairs which our method produced.

To further corroborate the results and in order to validate to what extent the similarity between their cuisine descriptions on Wikipedia might approximate the overall similarity between two cultures, we compare the similarity ranking of our communities with their corresponding ranking in the
cultural similarity index \cite{roose2010index} which is based on the European Social Survey (ESS).
ESS is a biennial 30-country survey of attitudes, beliefs and behaviour.  
This comparison reveals a low but significantly positive Spearman rank correlation ($\rho = 0.25, p < 0.001$). 

From our results we can conclude that: (1) the \emph{culinary similarities between communities produced by our approach seem to be plausible} (cuisines which have a very high ranking are perceived more similar than those with a very low ranking) and (2) \emph{shared internal states such as beliefs and values that may define a culture according to \cite{hofstede1980cultures,Alavi2004} are positively correlated with shared cultural practices that have been proposed as an alternative for defining a culture \cite{Bourdieu1984}}.
The relatively low correlation is of course not surprising since 
 ESS captures various variables (e.g., social and public trust; political interest and participation; socio-political orientations; media use; moral, political and social values), while our Wikipedia based approach focuses on culinary practices only. 
Including further cultural practices such as literature or art into our approach would most likely increase the correlation. 


\subsection{Cultural Understanding}

\begin{figure}[t!]
    \centering
    \setlength{\abovecaptionskip}{0ex}%
\setlength{\belowcaptionskip}{0ex}%
    \includegraphics[width=\linewidth]{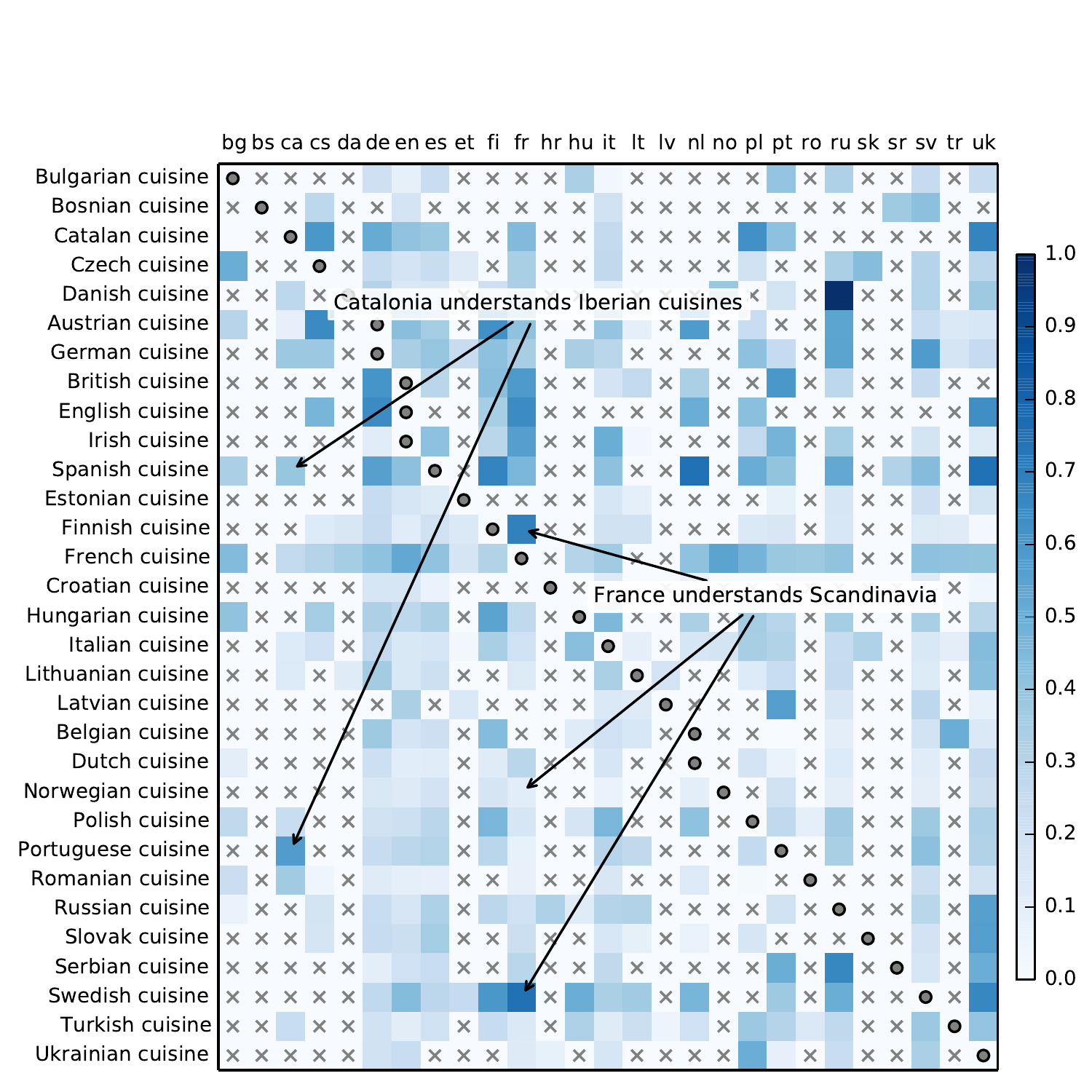}
    \caption[Heatmap of cultural understanding of cuisines on Wikipedia]{\textbf{Understanding of food cultures by different language communities:} The heatmap shows the overlap of concepts used to describe a cuisine by its associated language community and all other communities. The dots mark such associations (e.g. the Bulgarian Wikipedia community with the Bulgarian cuisine). The crosses represent missing data - i.e., cuisines which are not described by a certain language community. The absence of these articles might also be an indicator for (missing) cultural understanding.}
    \label{fig:04-cultural-understanding}
\end{figure}

To assess the cultural understanding of community A for the food culture of community B, we measure how similar community A describes the culinary practice of community B compared to how B describes it.
The understanding relation is asymmetric. 

\subsubsection{Results}
Figure \ref{fig:04-cultural-understanding} shows the cultural understanding for all community pairs. One can, e.g., see that the Catalan Wikipedia represents the Portuguese cuisine more accurately than the Spanish cuisine which might reflect the difficult relation between Catalonia and Spain. 
Also the good understanding of the Finnish and Swedish, but not the Norwegian cuisine by the French Wikipedia seems peculiar. 
Overall, the matrix is rather sparse, which could be interpreted as missing cultural understanding. If, for instance, the Hungarian Wikipedia does not have an article about the French cuisine, one could argue that the Hungarians are either not interested in cuisines in general or they are not interested in the French cuisine in particular. In both cases, this might be an indicator of missing cultural understanding, at least in the culinary domain. 



\begin{figure}[t!]
    \centering
    \includegraphics[trim=0 0 10 0,clip=true]{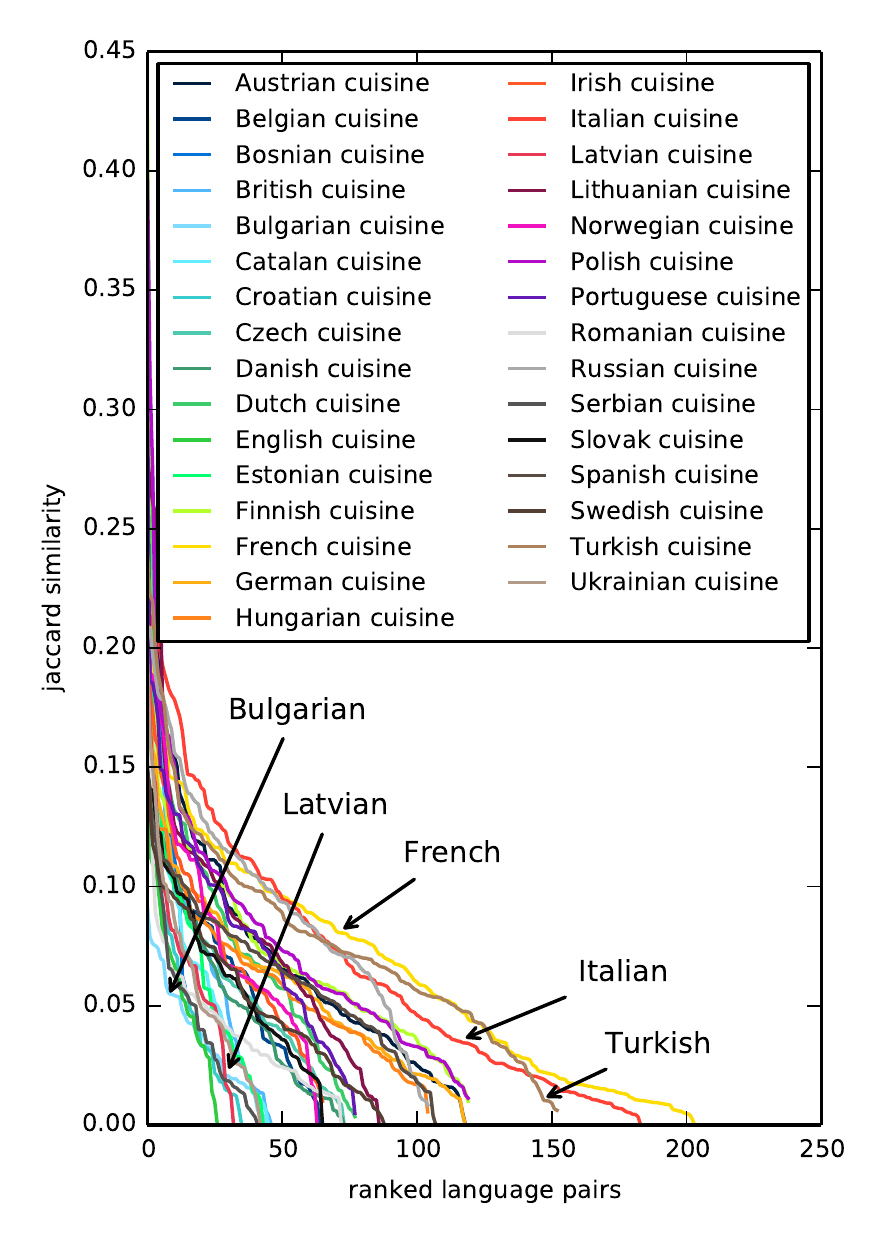}
    \caption[Ranked distribution of Cultural Similarities]{\textbf{Ranked distribution of cultural similarities:} 
    Each data point represents the jaccard similarity of two language-specific descriptions of one cuisine. The pairs are ranked by how similar they describe that cuisine from left (highest) to right (lowest) on the x-axis. 
    One can for example see that for the Bulgarian cuisine (light blue line), the two language editions which agree most on the description of that cuisine only reach a similarity of around 0.08. Popular cuisines such as the French and Italian cuisine are described by more language communities in a similar way while less popular cuisines, such as the Bulgarian or Latvian, are less similarly described by all communities.}

    \label{fig:03-01-topic-similarity}
\end{figure}

Apart from the question whether the internal perception of a cuisine differs from the external perception, it is also interesting to explore the variation of the perception of each food culture among different language communities. For all possible  combinations of language pairs, the overlap between the concepts used by the language communities to describe each cuisine is shown in Figure \ref{fig:03-01-topic-similarity}.
One can see that \emph{more prominent cuisines such as the Italian and French cuisine are more uniformly defined than less prominent cuisines such as the Bulgarian or Bosnian}. It seems noteworthy that the Turkish cuisine is described in a similar way by many language editions as well,  although it is not one of the most famous cuisines in Europe. A potential explanation for this observation is the high migration rate of the Turkish population in Europe. According to the bilateral migration database, they have the third highest negative migration after Russia and Poland.  

Figure \ref{fig:03-01-topic-similarity} also indicates that only a very small set of language communities define a food culture using the same concepts whereas the majority uses different ones. This supports the hypothesis that \emph{no globally true and objective description of culturally relevant practices exists on collaborative online production systems like Wikipedia, but descriptions are highly influenced by the cultural background of those who produce them}.

\begin{table}[b!]
\footnotesize
\centering
\begin{tabular}{l|ll}
\textbf{Pair}    & $\rho$ & ($p$-value) \\ \hline
wiki - ess       & $0.18$ & ($\ll 0.001$) \\
wiki - migration & $0.36$ & ($\ll 0.001$) \\
ess - migration  & $0.22$ & ($\ll 0.001$)
\end{tabular} 
\caption[Correlation results of cultural understanding with external data]{\textbf{Correlations of cultural understanding with external validation data:} The correlation values between the cultural understanding extracted by our approach (wiki), the values of an index derived from the European Social Survey (ess) and the migration data from the World Bank (migration). The cultural understanding between communities as inferred from Wikipedia can in smaller parts be explained by shared values, believes and behavior between those communities and in larger parts by migration. Migration manifests less in shared values, attitudes and behavior captured in the ESS index than in cultural understanding inferred from Wikipedia. }
\label{tab:04-cultural-understanding}
\end{table}

\subsubsection{Validation}
 Ideally, to validate the cultural understanding between community pairs in the culinary domain that we inferred from Wikipedia one would compare our results with external ground truth data. Since cultural understanding is a complex concept, no such universal ground truth data exists. However, it seems to be a plausible assumption that cultural understanding is at least partially influenced by cultural similarity (e.g., communities that are similar are likely to understand each other) and human mobility (e.g., countries that are the target destination of immigrants are assumed to understand the source country and culture of those immigrants rather well). 
For those influence factors external ground truth data exist, although not specifically related to cuisines.

We use the similarity index based on European Social Survey (ESS) \cite{roose2010index} as a rough approximation of cultural similarity between countries and data from the global bilateral migration database as a proxy for cross-country mobility patterns.
The lists of country pairs which were present in all three data sources (ESS, migration and wiki) were ranked as follows: The country pair AB with highest rank according to the migration statistics (Ukraine and Russia) is the pair where most people moved from country A to B. How many values and beliefs country A and B share determines the ranking of this pair in the ESS data. Finally, how well the language community associated with country A understands the food culture of country B  determines the ranking of this pair according to Wikipedia. 
Table \ref{tab:04-cultural-understanding} shows the Spearman rank correlations between these three ranked lists of pairs.
\emph{The cultural understanding which we inferred from Wikipedia seems to be better explained by understanding due to migration than due to shared values and beliefs}.
Further, migration explains to a lesser extent shared values and attitudes captured in the ESS index than cultural understanding inferred from Wikipedia.
This is a noteworthy finding, since it illustrates that \emph{the flow of people in Europe manifests more in shared understanding of culinary practices rather than shared values and beliefs}. 

\subsection{Cultural Affinity and Bias}
To assess the cultural affinity between two language communities we measure how much more attention one community pays to the food culture of the other one compared to how popular this food culture is from a global perspective.



\subsubsection{Results}
\begin{figure}[t!]
    \centering
    \captionsetup[subfloat]{justification=centering}
    \subfloat[]{\includegraphics[trim=24 0 0 0,clip=true,width=0.45\linewidth]{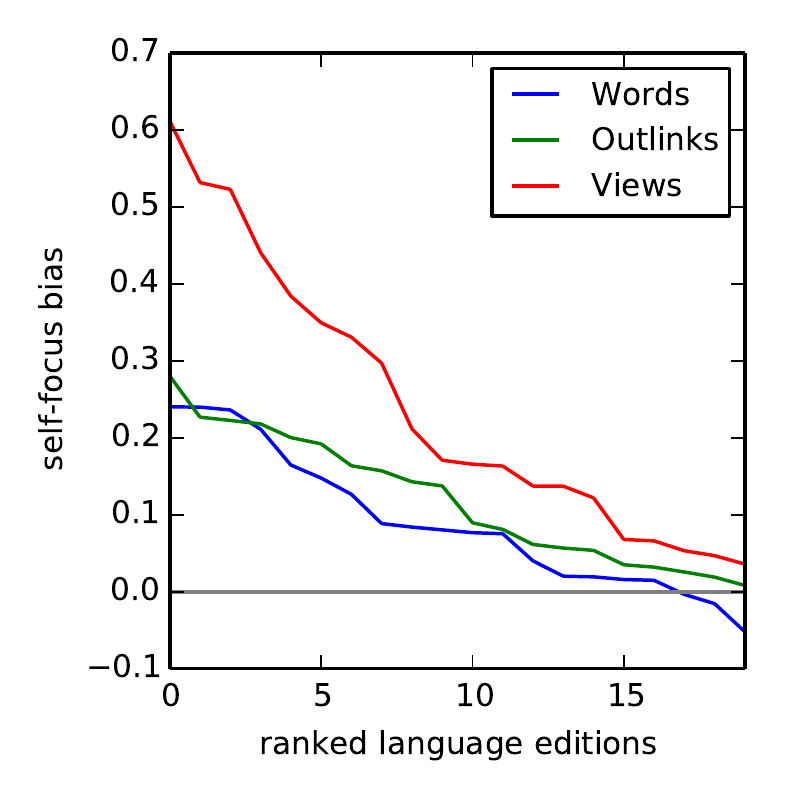}}
    \subfloat[]{\includegraphics[trim=24 0 0 0,clip=true,width=0.45\linewidth]{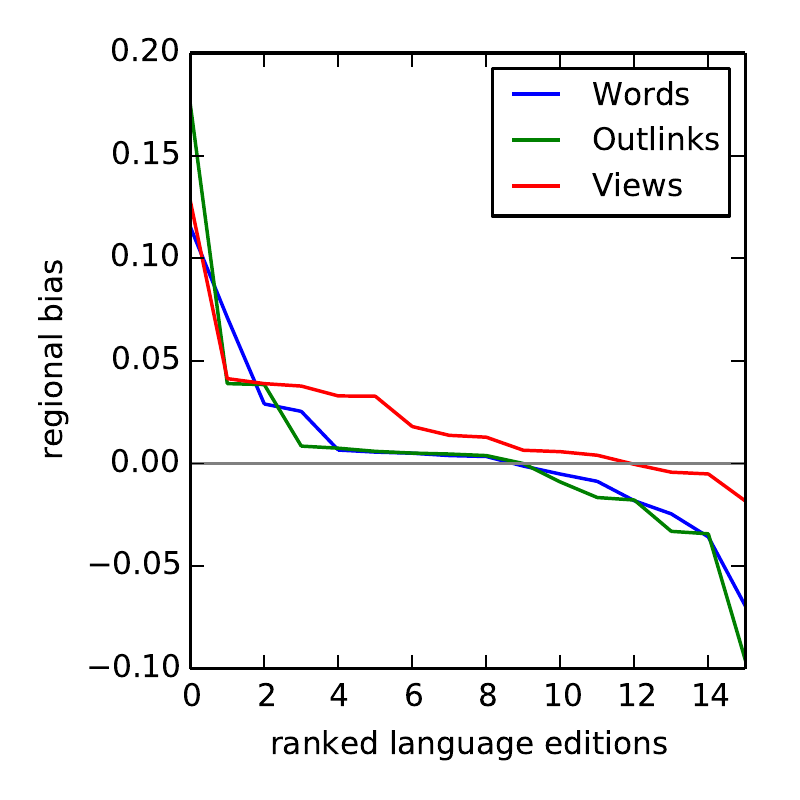}}
    \caption[Distribution of self-focus biases]{\textbf{Distribution of (a) self-focus and (b) regional biases:} The horizontal gray line indicates that no bias  is present -- i.e., foreign (or distant) cuisines are considered as important as the own (or neighbouring) cuisine. One cas see that most language editions reveal a strong positive bias towards their own cuisine, but only half of them show a slight positive tendency towards cuisines of neighbouring regions.}
    \label{fig:04-self-focus-regional-bias-distribution}
\end{figure}

 Previous work has shown that each Wikipedia edition tends to describe their locally relevant information in more detail than geographically distant concepts (cf. \cite{hecht2009measuring} \cite{overell2011world} \cite{maurer2006transformation}). Places like cities or monuments that are located in Finland are, for instance, described much more accurately and in greater detail on the Finnish Wikipedia than any other language edition \cite{hecht2009measuring}. 
Our results suggest that \emph{the self-focus phenomenon can also be observed for cuisines} (cf. Table \ref{tab:04-self-focus-regional-biases}).
However, it is interesting to note that we found lot of variation in the community-specific self-focus biases. Some countries and their corresponding communities reveal only small or moderate self-focus biases 
 while others like Bulgaria, Catalonia or Hungary seem to be especially interested in their own cuisines. Figure \ref{fig:04-self-focus-regional-bias-distribution} shows the distribution of the biases, where the self-focus bias is clearly visible. It also becomes apparent that \emph{the view data shows the highest self-focus bias, which means that consumers of Wikipedia are indeed much more interested in their own food culture, while editors seem to make an attempt to also represent relevant foreign food cultures}.



\begin{table}[b!]
\scriptsize
\centering
\begin{tabular}{l|r|r|r|r}
                  & \multicolumn{2}{c|}{Self-focus} & \multicolumn{2}{|c}{Regional focus}             \\
\textbf{Language} & \textbf{Outlinks} & \textbf{Views} & \textbf{Outlinks} & \textbf{Views} \\ \hline
Bulgarian         &     +0.227     &     +0.612     &       -0.033   &       -0.004   \\
Bosnian           &                &                &                &                \\
Catalan           &     +0.200     &     +0.523     &                &                \\
Czech             &     +0.062     &     +0.171     &       +0.000   &       +0.038   \\
Danish            &                &                &                &                \\
German            &     +0.054     &     +0.047     &       -0.018   &       -0.005   \\
English           &     +0.026     &     +0.036     &       -0.034   &       -0.000   \\
Spanish           &     +0.192     &     +0.137     &       +0.005   &       +0.041   \\
Estonian          &     +0.223     &     +0.384     &       +0.039   &       +0.018   \\
Finnish           &     +0.032     &     +0.163     &       +0.038   &       +0.033   \\
French            &     +0.138     &     +0.066     &       +0.008   &       +0.033   \\
Croatian          &                &                &                &                \\
Hungarian         &     +0.280     &     +0.441     &                &                \\
Italian           &     +0.035     &     +0.053     &       +0.006   &       +0.014   \\
Lithuanian        &     +0.019     &     +0.331     &       +0.005   &       +0.039   \\
Latvian           &                &                &                &                \\
Dutch             &     +0.057     &     +0.068     &       -0.096   &       -0.018   \\
Norwegian         &                &                &                &                \\
Polish            &     +0.143     &     +0.166     &       -0.009   &       +0.013   \\
Portuguese        &     +0.090     &     +0.122     &       +0.175   &       +0.128   \\
Romanian          &     +0.218     &     +0.532     &                &                \\
Russian           &     +0.008     &     +0.137     &       +0.007   &       +0.006   \\
Slovak            &                &                &                &                \\
Serbian           &                &                &                &                \\
Swedish           &     +0.157     &     +0.212     &       +0.004   &       +0.006   \\
Turkish           &     +0.081     &     +0.297     &                &                \\
Ukrainian         &     +0.164     &     +0.350     &       -0.017   &       +0.004   \\ \hline
Average           &     +0.120     &     +0.242     &       +0.005   &       +0.022   \\ \hline
Std.Dev           &      0.082     &      0.175     &        0.053   &        0.033   \\
\end{tabular}
\caption[Self-focus bias]{\textbf{Self-Focus Bias and Regional Bias:} The bias of a community towards their own and neighbouring cuisines ranges between -1 and +1. 
It becomes apparent that nearly all language communities on Wikipedia reveal a positive self-focus bias, while only half of them show a positive regional bias. Some values are missing since some Wikipedia editions describe less than three cuisines.}
\label{tab:04-self-focus-regional-biases}
\end{table}

In addition to the question whether a direct self-focus bias is visible, it is also a plausible assumption that geographic distance might impact affinities between communities \cite{spierdijk2006geography}. 
Table \ref{tab:04-self-focus-regional-biases} shows that the variance across communities with respect to their affinities and biases towards their neighbours is high. 
The Portuguese, Finnish and French Wikipedia communities seem to have stronger positive affinities towards their neighbouring communities.
Contrarily, the Netherlands, Bulgaria and the Ukraine do not show strong positive or negative affinities towards their neighbours.

Our results show that cross-cultural affinities on Wikipedia are distributed around zero but are slightly skewed to the right. That means, most communities do not have specific affinities towards each other, but some show much stronger positive affinities towards a foreign food culture than we would expect given the global popularity of that food culture. 

\subsubsection{Validation}
Assessing the validity of the cultural affinity relations for cuisines between communities which we extracted from Wikipedia is a difficult task since no obvious ground truth exists. Previous research suggests that the historical votes for the Eurovision Songcontest may expose politically motivated or culturally motivated affinities between communities \cite{spierdijk2006geography,garcia2013measuring} and might be used as a proxy for our use case.


Our results show that the view based affinities reveal the highest agreement with the affinities extracted from the Eurovision voting data (up to $0.25$ depending on the years covered). 
We further found that less affinity is exposed on Wikipedia compared to the Eurovision Songcontest voting data, which is not surprising since although different articles on Wikipedia compete for a limited amount of attention, there is no explicit competition going on like in the Eurovision Songcontest. 

\subsubsection{Simulation}
\begin{figure}[t!]
    \centering
    \subfloat[]{\includegraphics[width=0.5\linewidth]{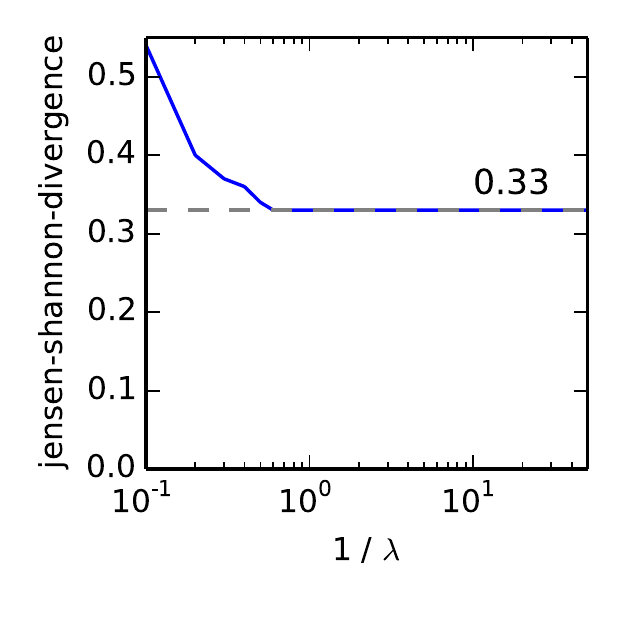}}
    \subfloat[]{\includegraphics[width=0.5\linewidth]{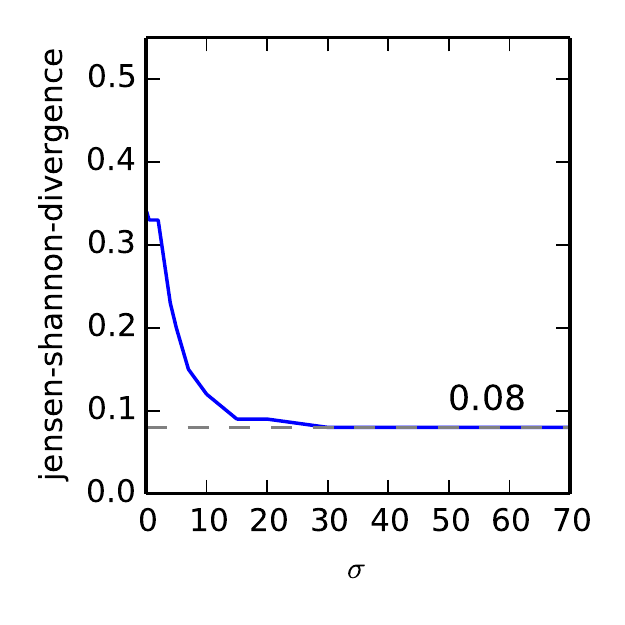}}
    \caption{\textbf{Jensen-Shannon (JS) divergence between the empirical data and two different models}. Model (a) corresponds to a global popularity model where only the global popularity of a practice influences how much attention it received. Popularity values are drawn from an exponential distribution with parameter $\lambda$. Model (a) can approximate the empirical data best if the popularity distribution is not too skewed and reaches a minimum divergence of $0.33$. Model (b) additionally considers community-pair specific affinities as an influence factor that may drive attention. The affinity values are drawn from a normal distribution with fixed mean ($\mu = 10.0$) and variance $\sigma$. One can see that Model (b) can approximate the empirical data much better than Model (a) and reaches a divergence of $0.08$ when the affinities between communities has a variance $\sigma \geq 30$.}
    \label{fig:jensen-shannon}
\end{figure}

To reveal the potential effect of affinities between communities on the observable outcome (the view data or the number of concepts per culturally relevant resource in different language editions), we simulate the process that might generate the data.
Our model receives as input a weighted network where all communities are connected. The weights are uniformly distributed if no affinity biases between communities exist (Model 1 in Figure \ref{fig:jensen-shannon} (a)) or can be drawn from a normal distribution (Model 2 in Figure \ref{fig:jensen-shannon} (b)) if we assume community-pair specific biases.
Further we assign to each community a popularity value which defines how popular the food culture of a community is - i.e., how much attention it will receive from all other communities independently of the presence of community-specific affinities.
Finally, each community may reveal a self-focus bias which is basically the affinity towards itself. In our simulations we set the self-focus bias to $0.242$ since our empirical results revealed that this is the average self-focus bias of a community on Wikipedia.


The first model (Figure \ref{fig:jensen-shannon} (a)) simulates a world where only the popularity of cultural practices determines how much attention they will receive from all other communities. The Jensen-Shannon-divergence (JS-divergence) with respect to the parameter $\lambda$ of the exponential function is shown in Figure \ref{fig:jensen-shannon} (a) and reveals how well the affinities of the simulated view data approximate our empirical observed affinity distribution. 
Our results show that if $\lambda$ is too big (i.e., the popularity distribution is too skewed) the empirical data is approximated worse than if we assume a less skewed popularity distribution. 
Nevertheless, the best simulation result shows a JS divergence of $0.33$. The JS-divergence would be zero if the two distributions were identical and approaches infinity the more they diverge.

The second model (Figure \ref{fig:jensen-shannon} (b)) uses the best popularity distribution (i.e., a popularity distribution which is not too skewed and therefore explains our empirical observations best) and draws affinities between communities from a normal distribution and different variance values. If the variance is zero, the second model is identical to the first model in the sense that affinity values are uniformly distributed. 
However, if the variance is greater zero, the model simulates a world where attention towards cultural practices is driven by the global importance of practices and cross-community specific affinities. The greater the variance the stronger the cross-community-specific affinities.

One can see in Figure \ref{fig:jensen-shannon} that a simulation model where affinities between community-pairs vary ($\sigma \geq$ 30) approximates the empirical data best (much better than the model based on popularity alone). 
This suggests that the process which generates the cross-cultural community attention on Wikipedia is driven by both factors, global popularity of cultural practices and cross-community specific affinities.

\section{Discussion}

In this paper we presented an approach for mining cultural relations from Wikipedia. In order to demonstrate the utility of our approach, we applied it to one specific cultural domain, cuisines and their representations on Wikipedia. Although our empirical results are limited to this domain, our approach is general and can be extended to further cultural dimensions, such as music, literature, arts and others.


 Keeping in mind the much broader scope of the external datasets (ESS, migration, song contest) used for empirical evaluation of our approach, the found correlations, in conjunction with the results from our crowdsourcing experiment and simulation are robust indicators that
 \emph{our approach indeed allows to extract meaningful cultural relations from Wikipedia}.

Our empirical results further show that \emph{communities that share more common values and beliefs according to the ESS index are also more likely to share culinary practices}.
This illustrates that a weak but significant correlation exists between shared internal states such as beliefs and values that may define a culture \cite{hofstede1980cultures,Alavi2004} and shared cultural practices \cite{Bourdieu1984}.

Not surprisingly, we found that \emph{prominent food cultures, such as the French and Italian cuisine are better understood} (i.e., their descriptions are more consistent across different language editions) than less prominent cuisines such as the Bosnian one. However, we also find that migration correlates with the cultural understanding between language communities as observed on Wikipedia to a certain extent. 
For example, besides the Italian and French cuisine the Turkish cuisine is among the three globally best understood cuisines and according to the bilateral migration database, Turkey has the third highest negative migration after Russia and Poland.  
Note that e.g. in Germany the largest part of the immigrant economy belongs to the food sector \cite{Mohring2008} and therefore migration is often related with the number of restaurants of foreign cuisine that exist in a country.

Related to the concept of cultural similarity and understanding is the concept of cultural affinity since one can hypothesize that communities which are e.g. more similar are also more interested in each other and vice versa. 
Our analysis of cross-cultural affinities on Wikipedia showed that affinities are distributed around zero but are slightly skewed. That means,\emph{ most communities do not have specific affinities towards each other, but some show much stronger positive bias towards a foreign food culture than we would expect} given the global popularity of that food culture. 
Remarkably, no strong negative affinities can be observed on Wikipedia - i.e. there are no communities which are significantly less interested in a foreign food culture that is popular from a global perspective. Our simulations confirm that \emph{affinities between communities are present} and allow us to reproduce the empirically observed cross-cultural view patterns much better than a model that assumes that the interest between communities only depends on the global importance of their cultural practices.
Finally, our results also show that \emph{all European communities reveal a substantial self-focus bias} (which corroborates related results from \cite{hecht2009measuring}) and \emph{a moderate bias towards geographically close region}s concerning cuisines (as in general also suggested by \cite{overell2011world,Zhu2013,Warncke-Wang2012}).


Analyzing the relation between cross-cultural affinities, understanding and similarities on Wikipedia suggests that a high understanding between two cultural distinct communities does not necessarily lead to positive affinities (Spearman $\rho$ between understanding and affinity is $-0.0393$), but high similarity tends to correlate positively with both, understanding and affinities (Spearman $\rho$ between similarity and understanding $+0.1880$ with $p \ll 0.001$ and between similarity and affinity $\rho=+0.287$ with $p \ll 0.001$). This suggests, that similar cultural groups (i.e., groups which have similar cultural practices) do not only understand each other, but also show high interest in each other, while dissimilar cultural groups that understand each other do not necessarily show high interest in each other.


\smallskip
\textbf{Implications:} 
We proposed a viable method to model and extract cultural relations from Wikipedia by analyzing cultural practices, on the example of culinary practices, on three distinct dimensions and showed how different language communities relate to each other in that domain. 
By extending the presented approach to domains such as music, literature, art, etc., it seems likely that it could become a powerful tool to complement reactive data collection such as surveys, which have been traditionally applied by social scientists and, e.g., market research firms to study how different cultural groups relate to each other.
Complementing reactive data collections for cross-cultural studies is an important issue since apart from other problems that notoriously plague survey-driven data collection, such as social desirability bias \cite{Meehl1946},  the  cultural background of the researcher may also introduce bias \cite{ailon2008mirror}.

\smallskip
\textbf{Limitations:} 
Language based comparisons of cultures are limited since language is only one aspect of culture and many different cultures and subcultures may share the same language. 
A widely used alternative is to rely on country borders.
However, cultures do not clearly map to countries since e.g. ethnic minorities may present a cultural group in a country which is different from the cultural group of the majority. This limitation is present in all current-state-of-art studies since no sources for cultural data at different aggregation levels exist so far.
Although many Wikipedia language editions can be associated with one or a small number of countries where the language is predominantly spoken, the community (i.e., the active as well as the passive part of the community) around a certain language edition of Wikipedia may not be representative for the population of this country. As a consequence, there is selection-bias present in Wikipedia data. However, sophisticated statistical methods exist that could in future work be applied to the data presented here to detect and correct such biases \cite{CUDDEBACK}.

\section{Conclusions}

Identifying biases and distorted descriptions of cultures on Wikipedia may help to guide the attention of the Wikipedia community towards areas where they need to invest effort (e.g., areas where the descriptions of cultural practices are biased and/or where very high or low cross-cultural interest exist and/or where cultural practices are extremely dissimilar). 
Our results suggest that nearly no unbiased descriptions of cultural practices in the national food domain exist on Wikipedia, leading to the suspicion that this might be similar for other cultural practices.
Unveiling those biases may on the one hand help to make people aware of them or correct them and may on the other hand reveal information about the relationship of different cultural communities.

\smallskip
\textbf{Main findings:}
The main empirical findings of this work are: (i) shared internal states such as beliefs and values that may define a culture are positively correlated with shared culinary practices, (ii) neighboring countries tend to have more similar cultural practices than more distant countries, (iii) cultural understanding can in part be explained by the global importance of a food culture (e.g., the French cuisine is ubiquitously better understood than the Bulgarian cuisine) and by migration (e.g., Turkish food culture is the third best understood food culture in Europe with an average understanding of $0.04$ across all language editions - after French and Italian food culture which both have an average understanding score of $0.06$ - and Turkey has the second highest emigration in Europe), (iv) almost all European language communities are most interested in their own cultural practice and only half of them show more interest in their neighbors' food culture than in the food culture of non-neighboring communities, (v) high understanding between two cultural distinct communities does not necessarily lead to high interest or vice versa, but high similarity tends to correlate positively with both, understanding and affinities. That means, similar cultural groups (i.e., groups which share cultural practices) do not only understand each other, but also show high mutual interest, while dissimilar cultural groups that understand each other do not necessarily show high interest in each other.

\smallskip
\textbf{Contributions:}
The contributions of this work are two-fold: (i) we present a novel approach for mining cultural relations between communities using the access volume and the content of the description of cultural practices on different language editions of Wikipedia and present its application and results based on the example of the culinary practices of 31 European countries; (ii) we validate our approach using several external datasets, crowdsourcing methods and simulations to show that the method is viable and merits further research and extension, as it shows much potential as a non-reactive way of empirically exploring cultural relations online.

\small
\bibliographystyle{abbrv}
\bibliography{food_biblio}

\begin{thebibliography}{10}

\bibitem{ailon2008mirror}
G.~Ailon.
\newblock Mirror, mirror on the wall: Culture's consequences in a value test of
  its own design.
\newblock {\em Academy of Management Review}, 33(4):885–904, 2008.

\bibitem{Alavi2004}
S.~B. Alavi and J.~McCormick.
\newblock Theoretical and measurement issues for studies of collective
  orientation in team contexts.
\newblock {\em Small Group Research}, 35:111--127, 2004.

\bibitem{bao2012omnipedia}
P.~Bao, B.~Hecht, S.~Carton, and M.~e.~a. Quaderi.
\newblock Omnipedia: Bridging the wikipedia language gap.
\newblock In {\em Proceedings of the SIGCHI Conference on Human Factors in
  Computing Systems}, CHI '12, pages 1075--1084, New York, NY, USA, 2012. ACM.

\bibitem{Bourdieu1977}
P.~Bourdieu.
\newblock {\em Outline of a Theory of Practice}.
\newblock Cambridge University press, 1977.

\bibitem{Bourdieu1984}
P.~Bourdieu.
\newblock {\em Distinction: A Social Critique of the Judgement of Taste}.
\newblock Harvard University Press, 1984.

\bibitem{callahan2011cultural}
E.~Callahan and S.~C. Herring.
\newblock Cultural bias in wikipedia content on famous persons.
\newblock {\em Journal of the Association for Information Science and
  Technology}, 62(10):1899--1915, 2011.

\bibitem{Calvo1982}
M.~Calvo.
\newblock Migration et alimentation.
\newblock {\em Social Science Information}, 21(3):383--446, 1982.

\bibitem{filatova2009multilingual}
E.~Filatova.
\newblock Multilingual wikipedia, summarization, and information
  trusworthyness.
\newblock {\em SIGIR workshop on information access in a multilingual world},
  2009.

\bibitem{garcia2013measuring}
D.~García and D.~Tanase.
\newblock Measuring cultural dynamics through the eurovision song contest.
\newblock {\em Advances in Complex Systems}, 16, 2013.

\bibitem{CUDDEBACK}
C.~Gary.
\newblock Detecting and statistically correcting sample selection bias.
\newblock {\em Journal of Social Service Research}, 3(30):19--33, 2004.

\bibitem{gavilanes2013cultural}
R.~O.~G. Gavilanes, D.~Quercia, and A.~Jaimes.
\newblock Cultural dimensions in twitter: Time, individualism and power.
\newblock In {\em International AAAI Conference on Weblogs and Social Media
  (ICWSM)}, 2013.

\bibitem{hecht2009measuring}
B.~Hecht and D.~Gergle.
\newblock Measuring self-focus bias in community-maintained knowledge
  repositories.
\newblock In J.~M. Carroll, editor, {\em Proceedings of the Fourth
  International Conference on Communities and Technologies}, pages 11--20. ACM,
  2009.

\bibitem{hecht2010tower}
B.~Hecht and D.~Gergle.
\newblock The tower of babel meets web 2.0: user-generated content and its
  applications in a multilingual context.
\newblock In {\em Conference on Human Factors in Computing Systems}, pages
  291--300. ACM, 2010.

\bibitem{hofstede1980cultures}
G.~H. Hofstede.
\newblock {\em Culture's consequences: International differences in
  work-related values}.
\newblock Sage Publications, Beverly Hills, CA, 1980.

\bibitem{Jermier1991}
J.~M. Jermier, J.~W. Slocum, L.~W. Fry, and J.~Gaines.
\newblock {Organizational Subcultures in a Soft Bureaucracy: Resistance behind
  the Myth and Facade of an Official Culture}.
\newblock {\em Organization Science}, 2(2):170--194, 1991.

\bibitem{Kayan2006}
S.~Kayan, S.~R. Fussell, and L.~D. Setlock.
\newblock Cultural differences in the use of instant messaging in asia and
  north america.
\newblock In {\em Proceedings of the 2006 20th Anniversary Conference on
  Computer Supported Cooperative Work}, CSCW '06, pages 525--528, New York, NY,
  USA, 2006. ACM.

\bibitem{Leidner2006}
D.~E. Leidner and T.~Kayworth.
\newblock Review: A review of culture in information systems research: Toward a
  theory of information technology culture conflict.
\newblock {\em MIS Q.}, 30(2):357--399, June 2006.

\bibitem{massa2013manypedia}
P.~Massa and F.~Scrinzi.
\newblock Manypedia: Comparing language points of view of wikipedia
  communities.
\newblock {\em First Monday}, 18(1), 2013.

\bibitem{maurer2006transformation}
H.~Maurer and J.~Kolbitsch.
\newblock The transformation of the web: How emerging communities shape the
  information we consume.
\newblock {\em Journal of Universal Computing Science}, 12(2):187--213, 2006.

\bibitem{Meehl1946}
P.~E. Meehl and S.~R. Hathaway.
\newblock The k factor as a suppressor variable in the minnesota multiphasic
  personality inventory.
\newblock {\em Journal of Applied Psychology}, 30(5):525, 1946.

\bibitem{Mohring2008}
M.~Möhring.
\newblock Transnational food migration and the internalization of food
  consumption: Ethnic cuisine in west germany.
\newblock {\em Food and Globalization: Consumption, Markets and Politics in the
  Modern World}, 1(1), 2008.

\bibitem{Michel2011}
J.-B. Michel and Y.~K. e.~a. Shen.
\newblock Quantitative analysis of culture using millions of digitized books.
\newblock {\em Science}, 331(6014):176--182, 2011.

\bibitem{nemoto2011cultural}
K.~Nemoto and P.~A. Gloor.
\newblock Analyzing cultural differences in collaborative innovation networks
  by analyzing editing behavior in different-language wikipedias.
\newblock {\em Procedia-Social and Behavioral Sciences}, 26:180--190, 2011.

\bibitem{overell2011world}
S.~E. Overell and S.~M. Rüger.
\newblock View of the world according to wikipedia: Are we all little
  steinbergs?
\newblock {\em Journal of Computational Science}, 2(3):193--197, 2011.

\bibitem{rask2008reach}
M.~Rask.
\newblock The reach and richness of wikipedia: Is wikinomics only for rich
  countries.
\newblock {\em First Monday}, 13(6), 2008.

\bibitem{roose2010index}
J.~Roose.
\newblock {\em Der Index kultureller Ähnlichkeit - Konstruktion und
  Diskussion}.
\newblock Freie Universität Berlin, 2010.

\bibitem{Schein1985}
E.~H. Schein.
\newblock {How Culture Forms, Develops, and Changes}.
\newblock In R.~H. Kilmann, M.~J. Saxton, and R.~Serpa, editors, {\em Gaining
  Control of the Corporate Culture}, pages 17--43+. Jossey Bass, 1985.

\bibitem{silva2014drink}
T.~H. Silva, P.~O. S.~V. de~Melo, J.~Almeida, M.~Musolesi, and A.~Loureiro.
\newblock You are what you eat (and drink): Identifying cultural boundaries by
  analyzing food and drink habits in foursquare.
\newblock In {\em International AAAI Conference on Weblogs and Social Media
  (ICWSM)}, 2014.

\bibitem{spierdijk2006geography}
L.~Spierdijk and M.~Vellekoop.
\newblock Geography, culture, and religion: Explaining the bias in eurovision
  song contest voting, 2006.

\bibitem{Talhelm2014}
T.~Talhelm, X.~Zhang, S.~Oishi, S.~Chen, D.~Duan, X.~Lan, and S.~Kitayama.
\newblock Large-scale psychological differences within china explained by rice
  versus wheat agriculture.
\newblock {\em Science}, 344(6184):603--608, 2014.

\bibitem{Tobler1970}
W.~Tobler.
\newblock A computer movie simulating urban growth in the detroit region.
\newblock {\em Economic Geography}, 46(2):234--240, 1970.

\bibitem{Warncke-Wang2012}
M.~Warncke-Wang, A.~Uduwage, Z.~Dong, and J.~Riedl.
\newblock In search of the ur-wikipedia: Universality, similarity, and
  translation in the wikipedia inter-language link network.
\newblock In {\em Proceedings of the Eighth Annual International Symposium on
  Wikis and Open Collaboration}, WikiSym '12, pages 20:1--20:10, New York, NY,
  USA, 2012. ACM.

\bibitem{West2004}
J.~West and J.~L. Graham.
\newblock A linguistic-based measure of cultural distance.
\newblock {\em MIR: Management International Review}, 44(3), 2004.

\bibitem{yanai2009detecting}
K.~Yanai, K.~Yaegashi, and B.~Qiu.
\newblock Detecting cultural differences using consumer-generated geotagged
  photos.
\newblock In {\em LocWeb}, ACM International Conference Proceeding Series,
  page~12. ACM, 2009.

\bibitem{Zhu2013}
Y.-X. Zhu, J.~Huang, and Z.-K. e.~a. Zhang.
\newblock Geography and similarity of regional cuisines in china.
\newblock {\em Computing Research Repository (CoRR)}, 2013.

\end{thebibliography}

\end{document}